\documentclass[aps,prd,groupedaddress,]{revtex4}
\usepackage{latexsym}
\usepackage[dvips]{graphicx}

\newcommand{\eq}{\begin{equation}}
\newcommand{\feq}{\end{equation}}
\newcommand{\eqn}{\begin{eqnarray}}
\newcommand{\feqn}{\end{eqnarray}}
\newcommand{\arr}{\begin{eqnarray*}}
\newcommand{\farr}{\end{eqnarray*}}
\newcommand{\beq}{\begin{equation}}
\newcommand{\eeq}{\end{equation}}
\newcommand{\bea}{\begin{eqnarray}}
\newcommand{\eea}{\end{eqnarray}}

\def\beq{\begin{equation}}
\def\eeq{\end{equation}}
\def\feq{\end{equation}}
\def\bea{\begin{eqnarray}}
\def\eea{\end{eqnarray}}
\def\bc{\begin{displaymath}}
\def\ec{\end{displaymath}}
\def\lb{\label}

\def\Ee{Entanglement entropy }
\def\ee{entanglement entropy }

\def\lb{\label}

\def\ord#1{O\left(#1\right)}

\begin{document}


\title{Holographic entanglement   entropy
of the BTZ black hole}

\author{Mariano Cadoni}
\email{mariano.cadoni@ca.infn.it}
\author{Maurizio Melis}
\email{maurizio.melis@ca.infn.it}
\affiliation{Dipartimento di Fisica,
Universit\`a di Cagliari, and INFN sezione di Cagliari, Cittadella
Universitaria 09042 Monserrato, ITALY}


\begin{abstract}

We  investigate quantum entanglement of 
gravitational configurations in 3D AdS gravity  using the AdS/CFT 
correspondence.   We derive  explicit formulas  for the 
holographic entanglement entropy (EE)
of the BTZ black hole, conical singularities and regularized AdS$_{3}$.  
The  leading term in the large temperature expansion of the holographic EE of the BTZ black hole 
reproduces exactly its 
Bekenstein-Hawking  entropy $S_{BH}$, whereas  the 
subleading term behaves as $\ln S_{BH}$.
We  also  show 
that the leading term  of
the  holographic EE for the BTZ black hole 
can be obtained from the 
large temperature  expansion of the partition function of a 
broad class of 2D CFTs on the torus.
This  result  indicates   that  black hole 
EE is not a fundamental feature of the underlying 
theory of 
quantum gravity but  
 emerges when the semiclassical notion of 
spacetime geometry is used to describe the black hole.

\end{abstract}


\maketitle
\section{Introduction}

At low energies any quantum theory of gravity must allow for  the
classical space-time description of general relativity.
Low-energy gravity is a
macroscopic phenomena that, at least to some extent, should be described
without  detailed knowledge of the fundamental microscopic theory
that holds at Planckian scales. From this point of view a
gravitational system is not very different from a condensed matter
system, whose macroscopical behavior allows for an effective
description in terms of low-energy degrees of freedom.
A strong evidence that this may work also  for gravity is represented
by the microscopic interpretation of the black hole entropy: in a
number of cases the Bekenstein-Hawking black hole entropy could be
reproduced as
Gibbs entropy, without detailed information about the underlying
microscopical description of quantum gravity degrees of freedom 
\cite{Strominger:1996sh,Horowitz:1996fn,Strominger:1997eq,
Cadoni:1998sg,Carlip:1998wz,
Carlip:2002be,Carlip:2006kq,Carlip:2007qh,Strominger:2009aj}.

A feature of many-body systems, which can be used to gain information
about macroscopic collective effects, is quantum entanglement. It
gives a measure of spatial correlations between parts of the system
and it is measured by the entanglement entropy (EE).
In the last years the notion of EE has been used with success as a 
tool for
understanding quantum phases of matter, but its application to
gravitational systems remains problematic \cite{Vidal:2002rm,Kitaev:2005dm,
Latorre:2004pk,korepin,
Casini:2004bw,Fursaev:2006ng,Fujita:2008zv,Casini:2009sr}.

The semiclassical  EE  of quantum
matter fields in a classical gravitational background (e.g. a black
hole) is not universal (it depends on the number of matter fields
species) and it is not clear if it can be extended to the quantum phase of
gravity \cite{'tHooft:1984re,Frolov:1993ym,Mann:1996ze}. The very notion of EE
for pure quantum gravity is not easy to define. The main obstruction comes
from the fact that in the usual Euclidean quantum gravity formulation
the metric, except its boundary value, cannot be fixed a priori (see e.g.  Ref. 
\cite{Fursaev:2007sg}),
whereas the usual, flat-space notion of EE requires to fix lengths in
bulk spacelike regions \footnote{A remarkable exception is represented 
by 2D AdS gravity. In 2D black hole entropy can be ascribed to 
quantum entanglement if Newton constant is wholly induced by quantum 
fluctuations \cite{Fiola:1994ir,Frolov:1996aj}. 
This fact allows a simple derivation of the EE of 2D 
AdS black hole \cite{Cadoni:2007vf,Cadoni:2007nh}.}.

A possible way out of these difficulties is to consider 
gravity theories with conformal field theory (CFT)
duals (see e.g. \cite{Aharony:1999ti}).
The advantage of considering   this kind of theories is 
twofold: 1) One can define the EE of a 
gravity configuration in terms of the EE of a field theory in 
which spacetime geometry is not dynamic;
2) At least for CFTs in 
two dimensions explicit and simple formulas for the EE are known 
\cite{Holzhey:1994we,Calabrese:2004eu,Calabrese:2009qy}.
The main drawback of this approach is related to the fact that the 
gravity/CFT correspondence is holographic (usually it takes the form 
of an AdS/CFT correspondence). Spatial correlations in the bulk 
gravity theory are codified in a highly nonlocal  way in the 
correlations of the boundary CFT.  This is particularly evident in 
the so-called UV/IR relation that relates large distances  on  
AdS space with the short distances behavior of the boundary CFT
\cite{Susskind:1998dq,Peet:1998wn}.

Because of this difficulty  the AdS/CFT correspondence has been only 
partially fruitful for understanding the EE of gravitational 
configurations, in particular of black holes. Some progress in this 
direction has been 
achieved  in the general case in Ref. \cite{Emparan:2006ni,
Solodukhin:2006xv,Hawking:2000da}
  and 
for the 2D case in Ref. \cite{Cadoni:2007vf,Cadoni:2007nh,Azeyanagi:2007bj}. 
Strangely enough, the AdS/CFT correspondence  has been used with much 
more success  in the reversed way, i.e. to 
compute the EE of boundary CFTs in terms of bulk geometrical
quantities \cite{Ryu:2006bv,Ryu:2006ef,Fursaev:2006ih,Hubeny:2007xt,
Michalogiorgakis:2008kk,Sun:2008uf,Nishioka:2009un}.

In this paper we will investigate quantum  entanglement in the 
context of 
three-dimensional (3D) AdS gravity, in particular  the 
Ba\~nados-Teitelboim-Zanelli (BTZ) black hole, using the AdS$_{3}$/CFT$_{2}$ 
correspondence.
We will tackle the problem using a standard method for studying 
correlations   in QFT: we will 
introduce in the boundary 2D CFT two external length-scales, a 
thermal wavelength $\beta=1/T$ ($T$ is the temperature of the CFT) 
and a  spatial length $\gamma$ which is the measure of the observable 
spatial region of our 2D universe. Varying $\beta$ we can probe 
thermal correlations of the CFT at different 
energy scales, whereas varying $\gamma$ we can probe the spatial 
correlations at different length scales.

We  will show that the AdS/CFT 
correspondence, and in particular the UV/IR relation, will allow us to 
identify in  natural way $\beta$ and $\gamma$ in terms of the two 
fundamental bulk length scales, the horizon 
of the BTZ black hole $r_{+}$ and the AdS length $L$.
This will allow us, using well-known formulas for the EE of 
2D CFTs and modular symmetry, to associate an ``holographic'' EE to
regularized 
AdS$_{3}$, the BTZ black hole and AdS$_{3}$ with conical  singularities.
We will also  show that the leading term in the EE of the BTZ black hole 
can be obtained in terms of the 
large temperature  expansion of the partition function of a 
broad class of CFTs on the torus. 
This  strongly supports  the intrinsic semiclassical nature of 
the black hole EE.

The structure of the paper is as follows.  We will 
briefly review some well-known facts  about the EE for 2D CFTs in Sect. II 
and about AdS$_{3}$ gravity  and the AdS$_{3}$/CFT$_{2}$ 
correspondence in Sect. III. In Sect. IV  we will discuss the modular 
invariance of the dual boundary CFTs on the torus.
In Sect. V we will investigate the relevance of  UV/IR relation for 
the calculation of the EE.
In sections VI, VII, VIII we will use our approach to derive the 
holographic  EE of, respectively, regularized AdS$_{3}$, the BTZ black hole 
and AdS$_{3}$ with conical singularities.
In section IX  we will compare  the holographic EE 
for the BTZ black hole with the large temperature  asymptotic 
expansion of the thermal entropy  of most common  2D CFTs on 
the torus. Finally, in Sect. X we will present our concluding remarks.

\section{Entanglement entropy of 2D CFT}
Most of the progress in understanding EE in QFT  has been achieved in 
the case of a 2D CFT. 
This is because the conformal symmetry can be used to determine the form 
of the correlation functions of the theory 
\cite{Holzhey:1994we,Calabrese:2004eu,Calabrese:2009qy}.

Let us consider a 2D spacetime with a compact spacelike dimension of
length $\Sigma$ and with $S^{1}$ topology.  When only a spacelike slice $Q$ (of length
$\gamma$) of our
universe is accessible for measurement, we loose information about the
degrees of freedom (DOF)  localized outside in the complementary
region $P$ and we have to
trace over these DOF.
The entanglement entropy   originated
by tracing over the unobservable DOF is given by the
von Neumann entropy
$S_{ent}=-
Tr_{Q}{\hat \rho}_{Q}\ln\hat\rho_{Q}$.
The reduced density matrix $\hat\rho_{Q}=Tr_{P}\hat\rho$ is obtained
by tracing
the density  matrix $\hat \rho$ over  states in the region $P$.

The resulting  EE for the ground state of the 2D CFT is given by
\cite{Holzhey:1994we,Calabrese:2004eu,Calabrese:2009qy}
\beq\lb{f5}
S_{ent}^{(\cal {C})}= \frac{c+\bar c}{6}\ln\left(\frac{\Sigma}{\epsilon
\pi}\sin\frac{\pi \gamma}{\Sigma}\right),
\feq
where $c$ and $\bar c$ are the central charges of the 2D CFT and 
$\epsilon$ is an ultraviolet cutoff necessary to regularize the 
divergence originated by  the  presence of a sharp boundary separating the region $P$
from the region $Q$.
Thus, Eq. (\ref{f5}) gives  the EE  for a   CFT at zero temperature 
and with a spacelike dimension with $S^{1}$ topology, i.e. for a 2D CFT  on a 
cylinder ${\cal {C}}$, whose timelike direction is  infinite.

For $\Sigma\gg \gamma$  the compact spacelike dimension becomes
also infinite and the EE is independent of 
$\Sigma$.  Eq. (\ref{f5}) gives the EE for
a 2D CFT at zero temperature on the plane ${\cal{P}}$:
\cite{Holzhey:1994we,Calabrese:2004eu,Calabrese:2009qy}
\beq\lb{f51}
S_{ent}^{(\cal {P})}= \frac{c+\bar c}{6}\ln\left(\frac{\gamma}{\epsilon}\right).
\feq
We can also consider a 2D CFT at finite temperature $T=1/\beta$ and a 
noncompact spacelike dimension. When only a spacelike slice of length
$\gamma$ is accessible to measurement,
the EE turns out to be that of a 2D CFT on a 
cylinder $C$, whose spacelike direction is  infinite
\cite{Calabrese:2004eu}:
\beq\lb{f52}
S_{ent}^{(C)}= \frac{c+\bar c}{6}\ln\left(\frac{\beta}{\epsilon
\pi}\sinh\frac{\pi \gamma}{\beta}\right).
\feq
It is important to stress that the cylinder $C$  can 
be obtained as the limiting case of a torus ${\cal {T}}(\beta,\gamma)$ with cycles 
of length $\beta$, $\gamma$, when $\gamma\gg\beta $. In Sect. IX 
we will use this 
feature  to relate the thermal entropy of a CFT on a torus with the 
EE of a CFT on the cylinder $C$.

\section {AdS$_{3}$ gravity and AdS$_{3}$/CFT$_{2}$ correspondence}
The EE of a QFT gives information about the spatial 
correlations of the theory. It follows that the EE of a 2D CFT, which  is the 
holographical 
dual of 3D gravity, should contain information about bulk
quantum gravity correlations. The most important  example in this context
is given by the correspondence between 3D AdS gravity and 2D CFT 
(AdS$_{3}$/CFT$_{2}$). 
Classical, pure AdS$_{3}$ gravity is  described by the  action
\beq\lb{e1}
A=\frac{1}{16\pi G_{3}}\int d^{3}x \left(R+ \frac{2}{L^{2}}\right),
\feq
where $L$ is the de Sitter length
and $G_{3}$ is 3D Newton constant.
The exact form of the 2D CFT dual 
to  3D AdS gravity  still remains a controversial point
\cite{Carlip:2005zn,Witten:2007kt,Maloney:2007ud}. However, in the 
large $N$ (central charge $c\gg 1$) regime, i.e. in region of 
validity of the gravity 
description, we know that the dual CFT  has  central charge \cite{Brown:1986nw}
\beq\lb{e2}
c=\bar c=\frac{3L}{2G_{3}}.
\feq
AdS$_{3}$ classical gravity allows for three kinds of  configurations. 
These  solutions of the action (\ref{e1})
can be classified in terms of orbits (elliptic, hyperbolic, parabolic)
of the $SL(2,R)$ group manifold \cite{Banados:1992wn,Banados:1992gq,Carlip:2005zn}.
The solutions corresponding to elliptic orbits  can be written
as
\beq\lb{e3a}
ds^{2}=  -\frac{1}{L^{2}}\left( r^{2}+
{r_{+}^{2}}\right)dt^{2}+
\left(r^{2}+r_{+}^{2}\right)^{-1}L^{2}dr^{2}+
\frac{r^{2}}{L^{2}}d\phi\, ,
\feq
where $0\le t\le\beta$, $0\le\phi\le 2\pi L$, $0\le r<\infty$ and 
$r_{+}$ is a constant.
The  corresponding 3D Euclidean space  has a contractible cycle
in the spatial,  $\phi$-direction .
For generic values of $r_{+}$ we have therefore a conical
singularity in this direction. Only for $r_{+}= L$ the
conical singularity disappears and the manifold becomes  nonsingular
3D AdS space at finite temperature $1/\beta$. The conformal boundary 
of the 3D spacetime is a torus with cycles of length $\beta$ and 
$2\pi L$. Correspondingly, the dual CFT will live in the torus ${\cal 
T}(\beta,2\pi L)$. The CFT on the 
cylinder $\cal C$ discussed in Sect. II  can be obtained in the limit 
$\beta \gg L$. This corresponds to  consider $-\infty <t <\infty$ and
$0\le\phi\le 2\pi L$.

The classical solutions of 3D gravity corresponding to
hyperbolic orbits of $SL(2,R)$ are
\beq\lb{e3}
ds^{2}= -\frac{1}{L^{2}} \left(r^{2}-r_{+}^{2}\right)dt^{2}+
\left(r^{2}-r_{+}^{2}\right)^{-1}{L^{2}}dr^{2}+
\frac{r^{2}}{L^{2}}d\phi^{2}.
\feq
Now the  3D Euclidean manifold  has a contractible cycle
in the  $t$-direction.
For generic values of $\beta$ and $r_{+}$  we have therefore a conical
singularity in this direction. Only for $\beta=\beta_{H}$, where
$\beta_{H}$ is the inverse Hawking temperature
\beq\lb{e4}
\quad \beta_{H}=\frac{1}{T_{H}}=
\frac{2\pi L^{2}}{r_{+}},
\feq
the conical singularity can be removed and the space describes
the Euclidean BTZ black hole.
The black hole has  horizon radius $r_{+}$, and mass and (thermal)
Bekenstein-Hawking (BH)  entropy given by
\beq\lb{tp}
M=\frac{r_{+}^{2}}{8G_{3} L^{2}},\quad S_{BH}= \frac{{\cal A}}{4G_{3}}=
\frac{\pi r_{+}}{2 G_{3}}.
\feq
Also in this case the conformal boundary 
of the 3D spacetime is the torus with cycles of length $\beta_{H}$, 
$2\pi L$ and the dual CFT will live on  
${\cal{T}}(\beta_{H},2\pi L)$. The CFT on the 
cylinder $C$   discussed in Sect. II  can be obtained in the limit 
$ L\gg \beta_{H}$. This corresponds to  consider a CFT at finite 
temperature, $0 \le t\le \beta_{H}$, with noncompact spacelike 
dimension
$-\infty<\phi< \infty$. In terms of the 3D bulk theory this 
corresponds to a  macroscopical black hole with $r_{+}\gg L$.

The separating element between the two classes of solutions above
corresponds to parabolic orbits of $SL(2,R)$,
\beq\lb{e4a}
ds^{2}= -\frac{1}{L^{2}} r^{2} dt^{2}+
\frac{L^{2}}{r^{2}}dr^{2}+
\frac{r^{2}}{L^{2}}d\phi^{2}, \quad -\infty <t<\infty.
\feq
The solution  can be seen as the $r_{+}=0$ ground state of the BTZ black hole,
i.e. the    $M=0,\, T_{H}=0$ solution. 

For $r_{+}\neq L$ the solution (\ref{e3a}) has a conical singularity
not shielded by an event
horizon \cite{Banados:1992wn,Banados:1992gq}. The conical singularity 
can also be  thought of as originated by a
pointlike source of mass $m$.
In the spectrum of AdS$_3$ gravity these solutions  are located 
between the NS vacuum, $r_{+}=L$, and the RR vacuum, $r_{+}=0$. Therefore we will
consistently take $0\le r_{+}\le L$.

Let us now briefly 
discuss the physical meaning of the conical singularity  
spacetime (\ref{e3a}).
To this end, let us rescale the coordinates in
Eq. (\ref{e3a}):
\beq\lb{rescaling}
r\to \frac{r_{+}}{L} r,\quad t\to \frac{L}{r_{+}} t,\quad \phi \to
\frac {L}{r_{+}} \phi.
\feq
The  metric becomes
\beq\lb{h1}
ds^{2}= -\left(\frac{r^{2}}{L^{2}}+1\right)dt^{2}+
\left(\frac{r^{2}}{L^{2}}+1\right)^{-1}dr^{2}+
\frac{r^{2}}{L^{2}}d\phi.
\feq
The previous expression describes  thermal AdS$_{3}$  in global
coordinates but,
owing to the rescaling of the coordinates we have now 
$0\le \phi\le 2\pi \Gamma L$, with
$\Gamma=r_{+}/L$.
The spacetime  has a conical singularity originated by a deficit
angle $2\pi(1-\Gamma)=2\pi(1- r_{+}/L)=2\pi(1-2\pi L/\beta_{con})$,
where  we have introduced 
\beq\lb{betacon}
\beta_{con}=2\pi L^{2}/r_{+},
\feq
as  the 
analogous of the inverse Hawking 
temperature $\beta_{H}$ to characterize the conical singularity.
In  the case of solution (\ref{e3}), setting
$\beta=\beta_{H}$   eliminates the conical singularity, whereas for 
solution (\ref{e3a})
we get a
regular manifold (AdS$_{3}$ at finite temperature)  for $\beta_{con}=2\pi L$.

The conical singularity we  have whenever $\beta_{con}\neq2\pi L$
represents
the geometric
distortion generated by a pointlike particle of mass
$m=(1-\Gamma)/4G_{3}$.
In order to find the holographic EE of the solution 
(\ref{e3a}), (\ref{e3}) and (\ref{e4a}), we have 
to  discuss first the modular symmetry of the  2D CFT dual to 3D AdS 
gravity.

\section{Modular Invariance}

It is well known that the partition function of a 2D CFT on the
complex torus
has to be invariant for transformation of the modular group $PSL(2,Z)$
\beq\lb{PSL}
\tau \to \frac{a \tau+b }{c\tau +d},
\feq
where $a,b,c,d$ are integers satisfying $ad-bc=1$,
$\tau=\omega_2/\omega_1$ is the modular parameter of
the torus and  $\omega_{1,2}$ are the  periods of the torus.
For simplicity we will take $\omega_{1}=\Sigma$ real and $\omega_{2}=
i\beta$ purely imaginary.
We are mainly interested in the modular transformation  of
the torus
\beq\lb{modt}
\tau \to -\frac{1}{\tau}.
\feq
3D spaces which are asymptotically
AdS are locally equivalent. The asymptotic form of the coordinate
transformations mapping the various spaces can be used to map one
into the other the tori describing the associated conformal
boundaries. For our discussion the relevant elements are the Euclidean BTZ black  hole
at Hawking temperature $1/\beta_{H}$,
AdS$_{3}$  space with  deficit angle $2\pi(1- 2\pi L/\beta_{con})$
 and  AdS$_3$ at
finite temperature $1/\beta_{H}$.
It will turn out that boundary tori associated with these three
spaces are related by modular transformations of the torus.

Let us  briefly review the well-known duality between the BTZ black hole
and AdS$_3$ at finite temperature \cite{Aharony:1999ti,Carlip:1994gc}.
To this purpose, we use the fact that the Euclidean BTZ  solution 
(\ref{e3})
with  periodicity $t\sim t+\beta_{H},\, \phi\sim \phi+2\pi L$ can
be mapped by a diffeomorphism into AdS$_3$ in Poincar\'e coordinates

\beq\lb{poincare}
 ds^2=\frac{1}{x^2}\left(dy^2+dzd\bar z\right),
\feq
where $z$ is a complex coordinate.

In the asymptotic $r\to \infty$ ($x\to 0$) region the map between the
BTZ black hole and AdS$_3$ in Poincar\'e coordinates is
\beq\lb{asmap}
z=\exp\left[\frac{2\pi}{\beta_{H}}\left(\phi+it\right)\right].
\feq
In order to have a natural periodicity, we introduce a new complex
variable $w$ 
\beq\lb{w}
z=\exp(-2\pi i w),
\feq
so that $w= (- t + i \phi)/\beta_{H}$.
One can now easily realize that the asymptotic conformal boundary of
the BTZ black hole is a complex torus with metric $ds^2= dwd\bar w$.
The periodicity of the imaginary  ($\omega_2$) and real ($\omega_1$)
part of $w$ are determined by the periodicity of $t,\,\phi$:
$\omega_2=2\pi i L/\beta_{H},\quad \omega_1=1$. The modular parameter
$\tau_{BTZ}=
\omega_2/\omega_1$ of the torus is therefore
\beq\lb{modp}
\tau_{BTZ}=\frac{2\pi i L}{\beta_{H}}.
\feq
Consider now Euclidean AdS$_3$ at finite temperature, described 
by the metric (\ref{h1}) with
the  periodicity
$t\sim t +\beta_{H}$ and $\phi\sim \phi+  2\pi L$. The $r\to \infty$
asymptotic form of  the map between AdS$_3$ at finite temperature and
AdS$_3$ in Poincar\'e coordinates is
\beq\lb{mapp}
z=\exp\frac{(t-i\phi)}{L},
\feq
whereas the coordinate $w$ of Eq. (\ref{w})
is now $w=\frac{1}{2\pi L}(\phi+it)$.
The  complex coordinate $w$ has now
periodicity $\omega_1=1,\,\omega_2=i\beta_{H}/2\pi L$. The
boundary of thermal AdS$_3$ is a torus with modular parameter
\beq\lb{modp1}
\tau_{AdS}=\frac{i\beta_{H}}{2\pi  L}.
\feq
Hence the boundary torus of the BTZ black hole and that of thermal
AdS$_3$ are related by the modular transformation
\beq\lb{modt1}
\tau_{AdS}=-\frac{1}{\tau_{BTZ}}.
\feq

Passing to consider the  Euclidean solution with the conical
singularity (\ref{e3a}),
we note that it is related to   AdS$_{3}$ just by the rescaling
(\ref{rescaling}). This changes the periodicity of the coordinates,
which becomes
$t\sim t+2\pi L$, $\phi\sim \phi+ 4\pi^{2}L^{2}/\beta_{con}$.
Because the coordinate transformation mapping
the boundary torus of conical singularity space into the boundary
torus of AdS$_{3}$ has the same form given by Eq. (\ref{mapp}),
it follows that the periodicity of the coordinate $w$ is
now $\omega_{1}= 2\pi L/\beta_{con},\, \omega_{2}=i$.
If we set $\beta_{con}=\beta_{H}$
the periodicity of the two tori are related by
\beq\lb{period}
\omega_{2}^{con}=\frac{i}{\omega_{1}^{AdS}},\quad
\omega_{1}^{con}=\frac{i}{\omega_{2}^{AdS}}.
\feq
The boundary torus of Euclidean AdS$_{3}$ with  conical singularity
characterized by the deficit angle $2\pi(1- 2\pi L/\beta_{H})$ has
the same modular parameter  as that of AdS$_{3}$ at temperature
$1/\beta_{H}$.
Notice that although
the two manifolds have the same topology and the same boundary torus,
they describe different three-geometries. The first is a singular
one, whereas the latter is a perfectly well-behaved geometry. For
this reason, one usually does not include AdS$_{3}$ with conical 
singularities  in the physical
spectrum of the theory.

Because $\tau_{con}=\tau_{AdS}$, from Eq. (\ref{modt1}) it follows
immediately  that, 
the boundary tori  of AdS$_{3}$ with conical
defect $2\pi(1- 2\pi L/\beta_{H})$ and that  of the BTZ black hole at
inverse
temperature $\beta_{H}$ are related by the modular transformation
\beq\lb{t1}
\tau_{con}=-\frac{1}{\tau_{BTZ}}.
\feq

\section{Entanglement entropy and the UV/IR relation}

As a consequence of the AdS/CFT correspondence  the 
EE (\ref{f5}), (\ref{f51}) and (\ref{f52}) should give 
information about bulk quantum gravity correlators.  More precisely, one 
would expect the EE in Eq. (\ref{f5}) to describe quantum correlations 
in the presence of conical singularity (\ref{e3a}) 
and the EE (\ref{f52}) of the thermal  CFT to describe the interplay 
between thermal and quantum correlations in the black hole 
background (\ref{e3}). The main obstacle to make the above relation 
precise is due to the holographic nature of the AdS/CFT correspondence.
Spatial correlations in the bulk gravity theory 
are codified in the boundary CFT in highly 
nonlocal way.  Whereas the inverse temperature $\beta$ appearing 
in Eq. (\ref{f52}) can be naturally identified as the inverse of the  black hole 
temperature (\ref{e4}),  the same is not true for the parameters $\gamma$ and 
$\epsilon$ in Eqs. (\ref{f5}),  (\ref{f51})  and  (\ref{f52}).

Owing to  the  holographic nature of the 
correspondence, the bulk interpretation  of these parameters requires
careful investigation. The AdS$_{p+1}$/CFT$_{p}$  correspondence 
indicates a way to relate length scales 
on the boundary with length scales on the bulk, this is  the UV/IR 
connection \cite{Susskind:1998dq,Peet:1998wn}. Infrared effects in  bulk, AdS$_{p+1}$ gravity   
correspond to ultraviolet  effects in the boundary CFT$_{p}$ and vice versa.
 
The UV/IR connection allows to identify the UV cutoff $\epsilon$ in 
Eq. (\ref{f51}) as an IR regulator of AdS$_{3}$ gravity 
\cite{Susskind:1998dq,Peet:1998wn}. This can be 
done in the usual way by using the dilatation isometry of the metric 
(\ref{e4a}) $r\to \lambda r$, $t\to \lambda^{-1} t$, $\phi\to 
\lambda^{-1} \phi$. Equivalently, one can  introduce ``cavity  coordinates'' on 
AdS$_{3}$ and show that $\epsilon$ acts as infrared regulator of the ``area'' 
of the S$^{1}$ boundary sphere \cite{Susskind:1998dq}. In fact, the regularized 
radius of the S$^{1}$ is $R=L^{2}/\epsilon$.  
The same is true in terms of the coordinate $r$ parametrizing  
AdS$_{3}$ in the modified Poincar\'e form (\ref{e4a}): cutting off at 
length scale $< \epsilon$ the 2D CFT implies an infrared  cutoff   
  on the radial coordinate of AdS$_{3}$, $r< \Lambda$ , where 
  \footnote{The length-scales  $\gamma$ and $\epsilon$ are defined 
  up to a dimensionless multiplicative constant of $\ord 1$.
  In the following we will  set this multiplicative constant equal 
  to $2\pi$.}

\beq\lb{IR}
\Lambda=\frac{4\pi^{2}L^{2}}{\epsilon}.
\feq 

The bulk interpretation of the parameter $\gamma$ in Eq. (\ref{f51}) 
is not as straightforward as that of $\epsilon$. $\gamma$ is not a 
simple external length scale we are using to cut off excitations of 
energy $< 1/\gamma$. It is the length of a {\it localized} spacelike 
slice of the  2D space on which the CFT lives. On the other hand, 
 owing to the holographic, nonlocal nature of the 
bulk/boundary correspondence, we expect that  any localization of DOF in the  
boundary  will be  lost by the correspondence with DOF on the bulk.
If any localization property  of the observable 
slice $Q$ is lost in the boundary/bulk duality, $\gamma$  can only 
play the role of an upper bound   above which spatial correlations for 
the boundary CFT are traced out. Because of the UV/IR connection, on 
AdS$_{3}$ this will correspond to  tracing out the bulk DOF at small 
values of the radial coordinate $r$, i.e. for 
$r<\omega$, where
\beq\lb{UV}
\omega=\frac{4 \pi^{2}L^{2}}{\gamma}.
\feq
It is important to stress  that the bulk parameter $\omega$ 
has not the same physical meaning of the boundary parameter 
$\gamma$.  Whereas $\gamma$  is the length of a spacelike  slice,
which is sharply separated from the observable region 
(hence it needs a UV regulator), $\omega$ has the much weaker meaning 
of a length scale below  which spatial correlations are traced out. 
In particular  in the AdS$_{3}$ bulk there is no sharp boundary 
separating observable and unobservable regions.

In the next sections we will use this meaning of $\gamma$ and 
$\omega$ to 
interpret the EE (\ref{f5}), (\ref{f51}) and (\ref{f52}) as 
holographic entanglement entropies of gravitational configurations.
 
\section{Holographic \Ee of regularized AdS$_{3}$ spacetime}
The AdS/CFT correspondence and the IR/UV connection described in the 
previous section allow us to   give to the EE (\ref{f51}) a simple 
bulk interpretation: it is the EE of regularized AdS 
spacetime (\ref{e4a}), i.e. it gives a measure of  the von Neumann entropy that arises 
when an IR cutoff $\Lambda$ is introduced and quantum gravity correlations 
are traced out for  $r<\omega$. Using Eqs. (\ref{IR}) and (\ref{UV}) 
into Eq. (\ref{f51}),
we have $S_{ent}^{AdS}= \frac{c}{3}\ln\left(\frac{\Lambda}{\omega}
\right)$ (we have used  $\bar c=c$).
The natural  length scale for  cutting off quantum bulk correlations is 
given by the AdS length $L$: $\omega=2\pi L$. This means that we are 
considering 
curvature effects much smaller than $1/L^{2}$. 
Using Eq. (\ref{UV}), this allows  the identification of the 
boundary parameter in terms of the AdS length $L$
\beq\lb{ID}
\gamma= 2\pi L.
\feq
The holographic EE of the regularized AdS spacetime 
\beq\lb{f51a}
S_{ent}^{AdS}= \frac{c}{3}\ln\left(\frac{\Lambda}{L}\right)
\feq
has a simple geometric 
interpretation. Apart from a proportionality factor,
it is the (regularized) proper length of the spacelike curve 
$t=const$, $\phi=const$. This can be easily shown integrating Eq.
(\ref{e4a}) for $L\leq r\leq \Lambda$.

\section {Holographic entanglement entropy of the BTZ black hole}
The spinless BTZ black hole (\ref{e3}) can be considered as
the thermalization at temperature 
$T=T_{H}$ of the AdS spacetime (\ref{e4a}). On the 2D boundary of the 
AdS spacetime, and  in the  above discussed large temperature limit 
$r_{+}\gg L$ , this  thermalization corresponds  to a plane/cylinder 
transformation that maps the  CFT on the plane ${\cal P}$ in the 
CFT on the 
cylinder $C$. The conformal map plane/cylinder has the (Euclidean) form
given in  Eq. (\ref{asmap}).
One can easily check that the above transformation is the asymptotic 
form of the map between the BTZ black hole and AdS$_{3}$ in Poincar\'e 
coordinates.
The conformal  transformation (\ref{asmap})  maps the EE of a 
CFT on the plane ${\cal P}$ in the EE of a CFT in the cylinder 
$C$ \cite{Calabrese:2004eu}, 
i.e. the EE of a CFT at zero temperature in a spacetime with 
noncompact  spacelike dimension into the EE of a CFT at finite 
temperature. As a result, Eq. (\ref{f51}) is transformed in Eq. 
(\ref{f52}) with $\beta=\beta_{H}$. Correspondingly, the holographic EE of the 
regularized AdS spacetime becomes the holographic 
EE of the BTZ black hole 
\beq\lb{e9}
S_{ent}^{{BTZ}}=S_{ent}^{CFT}(\gamma=2\pi L,\beta=\beta_{H})=
\frac{c}{3}\ln \frac{2 L^{2}}{ \epsilon
r_{+}}\sinh \frac{\pi
r_{+}}{L}.
\feq

The entanglement entropy (\ref{e9}) still depends on the UV cutoff
$\epsilon$. A renormalized entropy $\tilde S_{ent}^{{BTZ}}$ can be defined by
subtracting the contribution of the vacuum (the zero mass, zero
temperature BTZ  black hole solution).
In terms of the dual CFT we have to subtract the entanglement entropy
of the zero-temperature  vacuum state. 
This is given by  Eq. (\ref{f51}) with 
$\gamma=2\pi L.$
The renormalized entanglement entropy is therefore given by
\beq\lb{e11}
\tilde S_{ent}^{{BTZ}}=S_{ent}^{BTZ}-S_{ent}^{vac}= \frac{L}{2G_{3}} \ln
\frac{L}{\pi
r_{+}}\sinh \frac{\pi r_{+}}{L}.
\feq
As expected the renormalized entanglement entropy
vanishes for $r_{+}=0$ (the BTZ
black hole ground state).

The holographic entanglement entropy  (\ref{e11}) for the BTZ black hole
coincides exactly  with the previously derived entropy for the 2D AdS
black hole \cite{Cadoni:2007vf}. The 2D AdS black hole is the
dimensional reduction of the spinless BTZ black hole. 
Using the relationship between 2D and 3D
Newton constant $\Phi_{0}= L/4G_{3}¥$, Eq. (\ref{e11}) reproduces exactly
the result of Ref. \cite{Cadoni:2007vf}.

Macroscopic, i.e. large temperature, $r_{+}\gg L$, black holes  
correspond, in terms of the
2D CFT,  to the thermal wavelength $\beta_{H}$ much smaller
than the length  $2\pi L$. Expansion of Eq. (\ref{e11}) for 
$r_{+}/L\gg 1$ gives
\beq\lb{e12}
\tilde S_{BTZ}^{ent}= \frac{\pi r_{+}}{2 G_{3}¥}- \frac{L}{2G_{3}¥}\ln \frac{\pi
r_{+}}{L}+ O(1)= S_{BH}- \frac{L}{2G_{3}¥}\ln S_{BH}+ O(1).
\feq
The leading term  in 
entanglement entropy  is exactly the Bekenstein-Hawking entropy.
This leading term describes the extreme situation  
in which thermal fluctuations dominates completely. In this limit the
entanglement entropy  is just a measure of thermodynamical entropy. 
The EE (the von Neumann
entropy) for the  CFT  becomes extensive 
and it agrees with the Gibbs entropy of an
isolated system of length $\gamma= 2\pi L$. 
The subleading  term behaves as $\ln S_{BH}$ and describes the first corrections due
to quantum entanglement. 

The logarithmic correction in Eq. (\ref{e12}) matches exactly \footnote{Because the authors of Ref. 
\cite{Mann:1996ze} consider the case of a single scalar field, to 
reproduce their result we obviously need to use for the central 
charge $c+\bar c=1$} the short scale 
correction for the 
quantum entropy of  a scalar field in the BTZ Euclidean background found in 
Ref. \cite{Mann:1996ze}.
The fact that a 3D bulk calculation of the quantum entropy   give the 
same result of a boundary CFT calculation  is  not only a non  trivial 
check of their correctness, but  also a highly nontrivial check of the 
AdS$_{3}$/CFT$_{2}$ correspondence.

In principle, one could also consider the regime $\beta_{H}\sim
2\pi L$ in which the full quantum nature of the entanglement entropy should
be evident. However, this regime is singular from the black hole point of view: 
it  corresponds to the 3D analogous of the Hawking-Page
phase transition \cite{Hawking:1982dh,Kurita:2004yn}.

It is interesting to notice that the identification $\gamma=2\pi L$, 
which is crucial for deriving Eq. (\ref{e9}), can be
obtained 
without using the UV/IR connection, just assuming that 
 in the large $N$ 
limit the
mass/temperature relationship for the BTZ black hole exactly
reproduces that of a thermal 2D CFT. 

From Eqs. (\ref{e4}), (\ref{tp}) one easily finds the mass-temperature
relationship for
the BTZ black hole,
\beq\lb{e5}
M=\frac{\pi^{2}L^{2}}{2 G_{3}¥} T_{H}^{2}.
\feq
On the other hand, 
in the large temperature limit $\gamma\gg \beta$ the entanglement
entropy
(\ref{f52}) reduces to the classical, extensive thermal entropy
for an isolated system of length $\gamma$.
The energy/temperature relationship for such a 2D
CFT  is given by ($E_{0}$
is the energy of the vacuum)
 \beq\lb{e7}
 E-E_{0}= \frac{c }{12} \pi \gamma\left(T_{+}^{2}+ T_{-}^{2}\right),
 \feq
where $T_{+}$ and $T_{-}$ are the temperatures for the right and 
left oscillators. Identifying the black hole mass $M$ with $E-E_{0}$ and  the
 temperature $T_{H}=T_{+}=T_{-}$ of the CFT thermal state with the
 Hawking temperature of the black hole, we easily find, comparing Eq.
 (\ref{e7}) with Eq. (\ref{e5}) and using Eq. (\ref{e2}),
 $\gamma= 2\pi L$.

\subsection {Holographic Entanglement entropy of the rotating BTZ black hole}

The  derivation of the EE for the spinless BTZ black hole  
can be easily extended to the rotating BTZ
solution,

\beq\lb{f1}
ds^{2}= g(r)dt^{2}+
g(r)^{-1}dr^{2}+ r^{2}\left(\frac{d\phi}{L}-\frac{4 J G_{3}¥}{r}dt\right)^{2},\quad
g(r)=\frac{1}{r^{2}L^{2}}\left(r^{2}-r_{-}^{2}\right)
\left(r^{2}-r_{+}^{2}\right),
\feq
where $r_{\pm}$ are  the positions of outer and inner   horizons and
$J$ is the black hole angular momentum.
The thermodynamical parameters characterizing the black hole are the
mass  $M$, the angular momentum $J$, the Bekenstein-Hawking entropy $S_{BH}$, the
temperature $T_{H}$ and the angular velocity $\Omega$ (acting as
potential for $J$). These parameters satisfy the first principle
$dM=T_{H}¥dS_{BH}¥+\Omega dJ$ and can be written in terms of $r_{\pm}$:
\bea\lb{f2}
M&=&\frac{r_{+}^{2}+r_{-}^{2}}{8G_{3}¥L^{2}},
\quad J=\frac{r_{+}r_{-}}{4G_{3}¥L},\quad S_{BH}=\frac{\pi 
r_{+}}{2G_{3}¥}\, ,\\
T_{H}¥&=& \frac{1}{2\pi L^{2}} \left(
\frac{r_{+}^{2}-r_{-}^{2}}{r_{+}}\right),\quad \Omega=
\frac{1}{L}\frac{r_{-}}{r_{+}}\, .
\eea

The 2D CFT dual to the rotating BTZ black hole, although characterized
by the same central charge (\ref{e2}),  has different $L_{0}$
Virasoro
operators for the right and left movers. The eigenvalues of these 
operators corresponding to a BTZ black hole of mass $M$ and angular 
momentum $J$ are $L_{0}= 1/2(ML+J),\,
\bar L_{0}= 1/2(ML-J)$. The thermal density matrix for the CFT is given by
$\rho= \exp(-\beta H+ \beta \Omega P)$,  where $H$ and $P$ are the
Hamiltonian and the momentum operators.
In the canonical description of the thermal
2D CFT this amounts to consider  two different inverse temperatures
\beq\lb{f3}
\beta_{\pm}= \beta(1\pm \Omega)= 2\pi L^{2}(r_{+}\pm r_{-})^{-1}
\feq
for the right and left oscillators respectively.
The entanglement entropy for the thermal 2D CFT in the cylinder $C$ and for a spacelike slice of
length $\gamma$ is now given by \cite{Hubeny:2007xt}
\beq\lb{f4}
S_{ent}^{CFT}= \frac{c}{6}\ln\left[ \frac{\beta_{+}\beta_{+}}{\pi^{2}
\epsilon^{2}}
\sinh \frac{\pi \gamma
}{\beta_{+}}\sinh \frac{\pi \gamma
}{\beta_{-}}\right].
\feq
The length $\gamma$ can be determined in the same way as for the
spinless
BTZ black hole. For the thermal CFT with two different temperatures
for right and left movers we have the energy/temperature relation
\beq\lb{f5t}
E_{R}+E_{L}- E_{0R}-E_{0L}= \frac{c}{12} \pi\gamma\left(
T_{+}^{2}+T_{-}^{2}\right).
\feq
Using Eq. (\ref{f3}) into Eq. (\ref{f5t}) and comparing it with the black hole mass
(\ref{f2}), we obtain easily $\gamma=2 \pi L$.
As for the spinless case, we renormalize the entanglement entropy
by subtracting the contribution to the vacuum coming from the left
and right movers
$S^{ent}_{vac}= c/6(\ln(2\pi L/\epsilon)+\ln(2\pi L/\epsilon))$.
Putting all together, we get the renormalized entropy
\beq\lb{f5a}
\tilde S^{ent}_{BTZ}= \frac{L}{4 G_3}\ln\left[
\frac{L^{2}}{\pi^{2}(r_{+}+r_{-})(r_{+}-r_{-})}
\sinh \frac{\pi (r_{+}+r_{-})
}{L}\sinh \frac{\pi (r_{+}-r_{-})
}{L}\right].
\feq

Expanding the previous expression for $r_{+}\gg L$ and $r_{+}\gg r_{-}$
we get
\beq\lb{f6}
\tilde S^{ent}_{BTZ}= \frac{\pi }{2 G_3} r_{+} -\frac{L}{2 G_3} \ln \frac{ \pi
r_{+}}{2G_3} +O(1)= S_{BH}-\frac{L}{2 G_3} \ln S_{BH}+O(1).
\feq

\section{Holographic \ee of conical singularities}

Let us now consider the classical solution of 3D AdS gravity given by 
Eq. 
(\ref{e3a}), which describes 3D AdS spacetime with conical singularities.
As explained in Sect. IV, solution  (\ref{e3a}) can be locally obtained 
applying a diffeomorphism to the AdS spacetime (\ref{e4a}).  
This transformation is the ``spacelike'' counterpart of 
``thermalization''  mapping the  metric (\ref{e4a}) into the  
BTZ black hole. On the 2D conformal boundary of the 3D AdS spacetime 
this transformation is described by the map 
\beq\lb{mapp2}
z=\exp\frac{(t-i\phi)}{\beta},
\feq
where $\beta$ is easily determined by first applying the 
transformation (\ref{mapp}) mapping full AdS$_{3}$ into (\ref{e4a}) 
and then using the rescaling (\ref{rescaling}): $\beta=\beta_{con}$, 
where $\beta_{con}$ is given by Eq. (\ref{betacon}).
In the limit $\beta_{con}\gg 2\pi L$ (i.e. $L\gg r_{+}$)
the map (\ref{mapp2}) corresponds  to a plane/cylinder 
transformation that maps the  CFT on the plane ${\cal P}$ on the 
CFT on the cylinder $\cal{C}$.
Thus, this conformal  transformation  maps the EE of a 
CFT on the plane ${\cal P}$ in the EE of a CFT in the cylinder 
$\cal{C}$ \cite{Calabrese:2004eu}, 
i.e. the EE of a CFT at zero temperature  and noncompact spacelike 
dimension given by Eq. (\ref{f51})  into the EE of a CFT at zero  
temperature with a compact spacelike dimension given by Eq. 
(\ref{f5}). Correspondingly, the holographic EE of the 
regularized AdS spacetime becomes the holographic 
EE associated to AdS$_{3}$ with a conical singularity 
\beq\lb{e81ab}
S_{ent}^{con}=  \frac{c}{3}\ln\frac{\beta_{con}}{\pi \epsilon}
\sin \frac{2\pi^{2}L
}{\beta_{con}}=\frac{c}{3}\ln \frac{2L^{2}}{r_{+}\epsilon}\sin \frac{\pi
r_{+}}{L}.
\feq
Eq. (\ref{e81ab}) can be considered as the analytic continuation 
$r_{+}\to i r_{+}$ of Eq. (\ref{e9}). The holographic entanglement 
entropy of  a conical singularity described by a deficit angle 
$2\pi(1- 2\pi L/\beta_{con})$ is the analytic continuation of the 
holographic EE 
for the BTZ black hole with inverse temperature $\beta_{H}=\beta_{con}¥$.
The analytic continuation corresponds to the exchange of the 
(compact) timelike 
with the spacelike direction.   This result is a consequence of the 
modular symmetry (\ref{t1}) of the boundary CFT  on the torus 
relating the BTZ solution and the conical singularity spacetime. 
In the limit $r_{+}\gg L$ the boundary torus corresponding to the 
BTZ black hole can be approximated by the 
infinitely long (along the spacelike direction) cylinder $C$.
The modular transformation (\ref{t1})  maps the cylinder $C$ into the 
cylinder ${\cal C}$, which has infinitely long direction along the timelike 
direction and approximates the torus for $L\gg r_{+}$.
Correspondingly the EE for the BTZ black hole (\ref{e9}) is 
transformed in the EE for the conical singularity (\ref{e81ab}).

\section{\Ee of the BTZ black hole and thermal entropy of CFT on the torus}

In the previous sections we have discussed the holographic 
EE of gravitational configurations in 3D AdS spacetime.
In our approach the entanglement entropy of the boundary CFT,
$S_{ent}^{CFT}(\gamma,\beta)$, is used to probe thermal
correlations at scales set by $\beta$ and spatial correlations at
scales set by $\gamma$. The bulk description depends 
crucially on the regime of the AdS$_{3}$/CFT$_{2}$ correspondence 
we want to investigate. 

First of all, we work in the region of validity
of the gravity description of the AdS/CFT correspondence,  when the
AdS length is much larger than the Planck length,
\beq\lb{f9}
\frac{L}{G_{3}}\sim c\gg1,
\feq
that is in the large $N$ approximation.

Moreover, considering
curvature effects much smaller than the curvature of the AdS 
spacetime $1/L^{2}$ allows the identification of the external 
parameter $\gamma$ in terms of $L$. On the other hand, the thermal 
scale $\beta$  can be easily identified, when a black hole is present 
in the bulk: $\beta=\beta_{H}=1/ÇÈ T_{H}$. 
The semiclassical description for black holes
holds when the horizon radius is much
larger than the Planck length, $r_{+}\gg G_{3}$, whereas the holographic 
EE formula (\ref{e11}) holds for $r_{+}\gg L$.
We are in the 
regime where we are allowed to approximate the boundary torus with the 
cylinder $C$.  
 The path integral of Euclidean quantum gravity on AdS$_{3}$ is
dominated  by the contribution coming from the BTZ black hole at 
$T=T_{H}$.
The leading term in the EE (\ref{e12})  describes the main (thermal) 
contribution of the BTZ geometry and corresponds to 
the entanglement entropy for the CFT dominated completely by  thermal
correlations. 
When we increase the energy scale,  we reach a regime for which 
contributions coming from geometries  different from the BTZ instanton 
cannot be neglected.
Quantum entanglement and 
the subleading term in Eq. (\ref{e12})  become relevant.

The other regime we have investigated so far is $L\gg r_{+} $, which 
is related to the previous one by the modular transformation 
(\ref{t1}). The Euclidean quantum gravity partition function for 3D 
AdS gravity 
is now dominated by  AdS$_{3}$ at  temperature $T_{H}¥$. Although 
the solutions (\ref{e3a}) describe singular geometries with conical 
singularities - therefore they cannot be part of the physical 
spectrum of the theory - the modular symmetry strongly indicates that
they can be used  to probe quantum 
entanglement. In this regime  the boundary torus can be described by
the cylinder ${\cal 
C}$  and the EE is given by Eq. (\ref{e81ab}).

One may now wonder about the regime $r_{+}\sim L$. In this parameter region we 
cannot approximate the torus  with  an infinitely 
long cylinder.   $r_{+}=L$ is the fixed point of the modular 
transformations (\ref{modt1}), (\ref{t1}) and we have a  
large $N$ phase transition, which is the 3D analogue
of the  Hawking-Page transition \cite{Aharony:1999ti}.
Because now the dual boundary CFT  lives  in the torus 
${\cal T}(\beta_{H}¥,2\pi L)$, our calculations of the EE on the cylinder loose their 
validity. Furthermore, it is not a priori evident that the very 
notion of EE would maintain a sensible physical meaning in  a regime 
where the semiclassical description of gravity is expected to fail.

The most direct way to learn something about the 
relationship between the two regimes  $r_{+}\sim L$ and $r_{+}\gg L$ 
is to compare the $ L\gg \beta$ asymptotic behavior of the thermal 
entropy $S_{th} (\beta,L)$, derived from the partition 
function of the dual CFT on the torus, with the EE given by Eq.  
(\ref{e11}).
Unfortunately, whereas the EE for  a 2D CFT on a cylinder has an universal
form, the thermal entropy $S_{th}(\beta,L)$ for the CFT on the torus 
takes different form depending  on the details of the CFT we are dealing with 
\footnote{Despite the intense activity on
the subject in the last decade, the exact form of the 2D CFT dual 
to pure 3D AdS gravity  remains still a controversial point
\cite{Carlip:2005zn,Witten:2007kt,Maloney:2007ud}.}.

Here we will use a simple, albeit not completely general, approach to 
this problem. We will show that for the most common 2D CFTs (free 
bosons, free fermions, minimal models and Wess-Zumino-Witten models) 
the asymptotic, large temperature $L\gg \beta$ behavior of $S_{th}(\beta,L)$ 
calculated from the partition function of the CFT on the torus reproduces exactly 
the leading term of the EE (\ref{e11})  for the BTZ black hole.

The partition function of the CFT on the torus, $Z(\tau)$, is  a function of the 
modular parameter $\tau=i\beta/2\pi L$. 
Moreover, we will make use of the modular invariance of the partition 
function under the modular transformation (\ref{modt}) to write 
$Z(\tau)=Z(-1/\tau)$. From the partition function one can easily 
compute the thermal entropy
\beq\lb{te}
S_{th} = \log Z -\beta\partial_\beta (\log Z)\, .
\feq

We are interested in the asymptotic expansion 
of $S_{th}$ in terms of the variable 
\beq \lb{ipsilon}
y = \sinh\Big(\frac{2\pi^{2}¥ L}{\beta}\Big)\, ,
\feq
when  $y\to\infty$.
The asymptotic form of $S_{th}^{(as)}(y)$ is determined by first writing
$Z$ as a function of the usual  variable $q=\exp{(2\pi i\tau)}$.  
After making use of the modular invariance of  the partition function
under the modular 
transformation (\ref{modt}), we will introduce the new variable
$\tilde q=q(-1/\tau)=\exp{(-2\pi i/\tau)}$ and determine the $\tilde 
q\to 0$ asymptotic expansion of $Z(\tilde q)$. Finally, we will 
determine $S_{th}^{as}(y)$ by making use of the  $y\to \infty$ asymptotic 
expansion
\beq\lb{ase}
\tilde q= \frac{1}{4y^{2}}+{\cal O}\left(\frac{1}{y^{4}}\right).
\feq
Let us  sketch the results  of our calculations for the four 
cases under consideration.\\

\leftline{\sl Free bosons}
The partition function for free bosons on the torus is \cite{DiFrancesco:1997nk}
\begin{displaymath}
Z(\tau) = (\mbox{Im}\tau)^{-\frac{c}{2}}\,|\eta(\tau)|^{-2c}\, ,
\end{displaymath}
where $\eta$ is the  Dedekind function
\begin{displaymath}
\eta(\tau) = q^{\frac{1}{24}}\prod_{n=1}^\infty(1-q^n).
\end{displaymath}
The $\tilde q\to 0$ asymptotic expansion for  the entropy turns 
out to be
\beq\lb{fb}
S_{th}(\tilde q)=-\frac{c}{6}\ln{\tilde q}+{\cal 
O}\left(\ln(-\ln\tilde q)\right),
\feq
whereas the $y\to\infty$ expansion is 
\beq\lb{fb1}
S_{th}(y)=\frac{c}{3}\log{y}+{\cal 
O}\left(\ln(\ln y)\right)\, .
\feq

\leftline{\sl Free fermions}
The partition function for free fermions on the torus is \cite{DiFrancesco:1997nk}
\begin{displaymath}\lb{ff}
Z(\tau) = \sum_{i=2}^4\left|\frac{\theta_i(\tau)}{\eta(\tau)}\right|^{2c}\, ,
\end{displaymath}
where we have introduced the modular functions:
\begin{eqnarray}
\theta_2(\tau) & = & 2q^{\frac{1}{8}}\prod_{n=1}^\infty(1-q^n)(1+q^n)^2\, 
,\nonumber \\
\theta_3(\tau) & = & \prod_{n=1}^\infty(1-q^n)(1+q^{n-\frac{1}{2}})^2\,
,\nonumber \\
\theta_4(\tau) & = & \prod_{n=1}^\infty (1-q^n)(1-q^{n-\frac{1}{2}})^2\, .
\nonumber
\end{eqnarray}
In this case we find the  $\tilde q\to 0$ and $y\to \infty$ asymptotic 
expansion for  the entropy 
\beq\lb{fb2}
S_{th}=-\frac{c}{6}\ln{\tilde q}+{\cal O}(1)=\frac{c}{3}\ln{y}+{\cal O}(1).
\feq

\leftline{\sl Minimal models}
The partition function now is given  by \cite{DiFrancesco:1997nk}
\begin{displaymath}\lb{mm}
Z(\tau) = \sum_{h,\,\bar h}\chi_{_h}(\tau){\cal M}_{_{h,\,\bar
h}}\,{\bar\chi}_{_{\bar h}}
(\bar\tau)\, ,
\end{displaymath}
where $\chi_{h}$ are the Virasoro characters:
\begin{displaymath}\lb{vc}
\chi_{_h}(\tau) = \frac{q^{h-\frac{c-1}{24}}}{\eta(q)}\, .
\end{displaymath}
and ${\cal M}_{_{h,\,\bar
h}}$ are the so-called mass matrix elements.
Using the asymptotic form for the Euler $\varphi$-function,
\beq\lb{ef}
\varphi({\tilde q}) = (1-{\tilde q})\big[1+{\cal O}({\tilde q}^2)\big]\,,
\feq
one finds that the partition function has the following $\tilde q\to 0$ asymptotic 
expansion
\beq\lb{mm1}
Z = A\, {\tilde q}^{-\frac{c}{12}}\,(1-{\tilde q})^{\alpha}
\big[1+{\cal O}({\tilde q}^2)\big]\, ,
\feq
where  $A = {\cal M}_{_{0,\,0}}$, $\alpha = -2(d+1)$ and
$d={\cal M}_{_{1,\,0}}/{\cal M}_{_{0,\,0}}$.

Using Eqs. (\ref{te}) and (\ref{ipsilon}) one finds also in this case 
for the entropy the same asymptotic form given in Eq. (\ref{fb2}).

\smallskip

\leftline{\sl Wess-Zumino-Witten models}
The partition function for Wess-Zumino-Witten models is \cite{DiFrancesco:1997nk}
\begin{displaymath}\lb{wzw}
Z(\tau) = \sum_{\hat\lambda,\,\hat\xi}\chi_{_{\hat\lambda}}(\tau)
{\cal M}_{_{\hat\lambda,\,\hat\xi}}\,{\bar\chi}_{_{\hat\xi}}(\bar\tau)\, ,
\end{displaymath}
with the characters  $\chi_{_{\hat\lambda}}$ given by:
\begin{displaymath}\lb{wzw1}
\chi_{_{\hat\lambda}}(\tau)\equiv \chi_{_{\lambda_1}}^{(k)}(\tau) = 
\frac{q^{\frac{(\lambda_1+1)^2}{4(k+2)}}}{\big[\eta(q)\big]^3}\;
\sum_{n\in  Z}\big[\lambda_1+1+2n(k+2)\big]q^{n[\lambda_1+1+(k+2)n]}\, .
\end{displaymath}
The asymptotic expansion for the characters is
\begin{displaymath}
\chi_{_{\lambda_1}}^{(k)}(-1/\tau) = 
\frac{{\tilde q}^{\frac{(\lambda_1+1)^2}{4(k+2)}-\frac{1}{8}}}
{\big[\varphi({\tilde q})\big]^3}\,\left[\lambda_1+1+
{\cal O}\big({\tilde q}^{k-\lambda_1+1}\big)\right]\, ,
\end{displaymath}
from which it follows the ${\tilde q}\to 0$ asymptotic form of the 
partition function: 
\begin{eqnarray}
Z & = & \big[\varphi({\tilde q})\big]^{-3}
\sum_{\lambda_1,\,\mu_1}{\cal M}_{_{\lambda_1,\,\mu_1}}
{\tilde q}^{\frac{(\lambda_1+1)^2+(\mu_1+1)^2}{4(k+2)}-\frac{1}{4}}\times \nonumber \\
& & \left[\lambda_1+1+{\cal O}\big({\tilde q}^{k-\lambda_1+1}\big)\right]\times
\left[\mu_1+1+{\cal O}\big({\tilde q}^{k-\mu_1+1}\big)\right]\, .\nonumber
\end{eqnarray}
The leading terms in the previous summation are those with $\lambda_1 
= \mu_1 =0$, so that, using the asymptotic form (\ref{ef}) for the 
$\varphi$-function, one finds
\beq\lb{gg1}
Z = {\cal M}_{_{0,\,0}}\,{\tilde q}^{-\frac{k}{4(k+2)}}\,
(1-{\tilde q})^{-3}\big[1+{\cal O}\big({\tilde q}^2\big)\big]\, .  
\feq
Specializing to the case of an $su(2)_{_k}$ algebra, for which one 
has $c = \frac{3k}{k+2}$, Eq. (\ref{gg1}) takes the form given by Eq.
(\ref{mm1}) with $A = {\cal M}_{_{0,\, 0}} \;\; 
\mbox{and} \quad \alpha = -3\,$.
Both $A$ and $\alpha$ do not enter in the leading term of the 
asymptotic expansion for the entropy, but determine only  the 
subleading terms. 
Thus, making use of Eqs. (\ref{te}) and (\ref{ipsilon}), one finds also in this case 
for the entropy the same asymptotic form given in Eq. (\ref{fb2}).

\section{Final remarks}
Let us first summarize the main result of the previous section. The leading term in 
the large temperature, $y\to\infty$ expansion of the thermal entropy 
for the four CFT classes on the torus considered in this paper,
\beq\lb{cc}
S_{th}\sim\frac{c}{3}\ln{y}=\frac{c}{3}\ln\sinh\frac{2\pi^{2}L}{\beta},
\feq
reproduces 
for $\beta=\beta_{H}$ the leading term of the  holographic EE for the BTZ black hole given by Eq. 
(\ref{e11}). This  result  sheds light on the 
meaning of the holographic EE for the BTZ black hole in 
particular and, more in general, on the very meaning of entanglement 
for black holes. In fact our result  indicates that 
entanglement entropy for black hole is a semiclassical concept that 
has a meaning only for macroscopical black holes in the regime $r_{+}¥\gg L$.
Thus,  entanglement  seems to arise from a purely thermal description of the 
underlying quantum  theory of gravity 
which is assumed to describe 3D quantum gravity in 
the region $r_{+}\sim L$. 
This fact supports the point of view that the microscopic theory 
describing the BTZ black hole at short scales is unitary. 
Entanglement entropy is an 
emergent concept, which comes out when the semiclassical notion of 
spacetime geometry is used to describe the black hole.
The agreement between thermal entropy for the CFT on the torus and 
holographic EE for the BTZ black hole is limited to the leading term 
in the $y \to \infty $ expansion. 
The subleading terms in the expansions (\ref{fb1}) and (\ref{fb2}) are 
not of the same  order for the different CFT we have considered. The 
subleading terms are of order $\ln(\ln y)$ for the  free boson, 
whereas they are $\ord 1$ for the other three cases. These subleading 
terms  seem to be not universal but they depend on the actual CFT we are 
dealing with.

An other important point, which we have only partially addressed in this paper,
concerns the role played by the classical solutions of 
3D AdS gravity describing conical singularities of the spacetime.
Because they represent singular geometries, they cannot be part of the 
physical spectrum of pure 3D AdS gravity (although they may play a 
role for gravity interacting with pointlike matter).
On the other hand,  they are related with the BTZ black hole solutions 
by modular transformations and one can associate to them an 
entanglement entropy. 
All this  could be very useful for shedding light on the phase 
transition (analogue to the Hawking-Page transition of 
four-dimensional gravity), which is expected to take place at 
$r_{+}=L$.

\end{document}